\documentclass[11pt, letterpaper, logo]{appier-ai-research}

\PassOptionsToPackage{comma,numbers,sort,compress}{natbib}

\usepackage[comma,numbers,sort,compress]{natbib}

\usepackage{hyperref}[citecolor=magenta]
\usepackage{fdsymbol}

\hypersetup{
    colorlinks = true,
    citecolor = {magenta},
}

\pdftrailerid{redacted}

\makeatletter
\renewcommand\bibentry[1]{\nocite{#1}{\frenchspacing\@nameuse{BR@r@#1\@extra@b@citeb}}}
\makeatother

\usepackage{kantlipsum, lipsum}
\usepackage{dsfont}
\usepackage{gdm-colors}
\usepackage{subfigure}
\usepackage{bbm}
\usepackage{wrapfig}
\usepackage{multirow}
\usepackage{amsmath}
\usepackage{amssymb}
\usepackage{graphicx}
\usepackage{paralist}
\usepackage{enumitem}

\usepackage[most,skins,theorems]{tcolorbox}

\tcbset{
  aibox/.style={
    width=\linewidth,
    top=8pt,
    bottom=4pt,
    colback=blue!6!white,
    colframe=black,
    colbacktitle=black,
    enhanced,
    center,
    attach boxed title to top left={yshift=-0.1in,xshift=0.15in},
    boxed title style={boxrule=0pt,colframe=white,},
  }
}
\newtcolorbox{AIbox}[2][]{aibox,title=#2,#1}
\definecolor{lightblue}{rgb}{0.22,0.45,0.70}

\tcbset{  
  aiboxc/.style={
    width=\columnwidth,
    colback=blue!6!white,
    colframe=black,
    colbacktitle=black,
    enhanced,
    center,
    attach boxed title to top left={yshift=-0.1in,xshift=0.15in},
    boxed title style={boxrule=0pt,colframe=white,},
  }
}
\newtcolorbox{AIboxC}[2][]{aiboxc,title=#2,#1}

\graphicspath{{figures/}}

\title{None of the Above, Less of the Right\\Parallel Patterns between Humans and LLMs on Multi-Choice Questions Answering}

\correspondingauthor{ray.tam@appier.com, brian.wu@appier.com}

\author[1]{Zhi Rui Tam}
\author[1]{Cheng-Kuang Wu}
\author[1]{Chieh-Yen Lin}
\author[2]{Yun-Nung Chen}

\affil[1]{Appier AI Research}
\affil[2]{National Taiwan University}
\vspace{-0.5cm}

\begin{abstract}
\vspace{-0.4cm}
Multiple-choice exam questions with "None of the above" (NA) options have been extensively studied in educational testing, in which existing research suggests that they better assess true knowledge.
However, their impact on Large Language Models (LLMs) evaluation remains underexplored.
Through systematic experiments with 28 LLMs on the MMLU benchmark, we examine how NA options affect model performance and confidence calibration.
Our analysis reveals that NA options, when used as the correct answer, lead to a consistent 30-50\% performance drop across models regardless of scale--suggesting that LLMs lack the meta-cognitive ability to systematically evaluate and reject all given options when none are correct.
This degradation shows strong domain dependence, with minimal impact on mathematical reasoning (14.6\% drop) but severe effects on tasks requiring uncertainty handling like business ethics (48.1\% drop).
Our results highlight important implications for benchmark design and raise questions about LLMs' ability to handle uncertainty in real-world applications.
\end{abstract}

\begin{document}

\maketitle

\newcommand{\expect}[2]{\mathds{E}_{{#1}} \left[ {#2} \right]}
\newcommand{\myvec}[1]{\boldsymbol{#1}}
\newcommand{\myvecsym}[1]{\boldsymbol{#1}}
\newcommand{\vx}{\myvec{x}}
\newcommand{\vy}{\myvec{y}}
\newcommand{\vz}{\myvec{z}}
\newcommand{\vtheta}{\myvecsym{\theta}}

\begin{abstract}
\end{abstract}

\vspace{-0.25cm}
\section{Introduction}
\vspace{-0.2cm}
\label{intro}

Multiple-choice question answering (MCQA) benchmarks—such as MMLU \citep{hendrycks2020measuring} and MMLU-Pro \citep{Wang2024MMLUProAM}—have become a cornerstone for evaluating large language models (LLMs) by measuring their domain-specific knowledge and reasoning capabilities.
Originally designed for human educational assessments, these benchmarks adhere to well-established guidelines for item construction and distractor design (e.g. \citep{haladyna2002review, piontek2008best}).
Yet a critical gap persists: guidelines developed for human test situations, which can enhance both reliability and discrimination \citep{rich1990item}, are rarely scrutinized in the context of LLM evaluation.

\begin{figure}[t!]
  \centering
  \includegraphics[width=0.5\columnwidth]{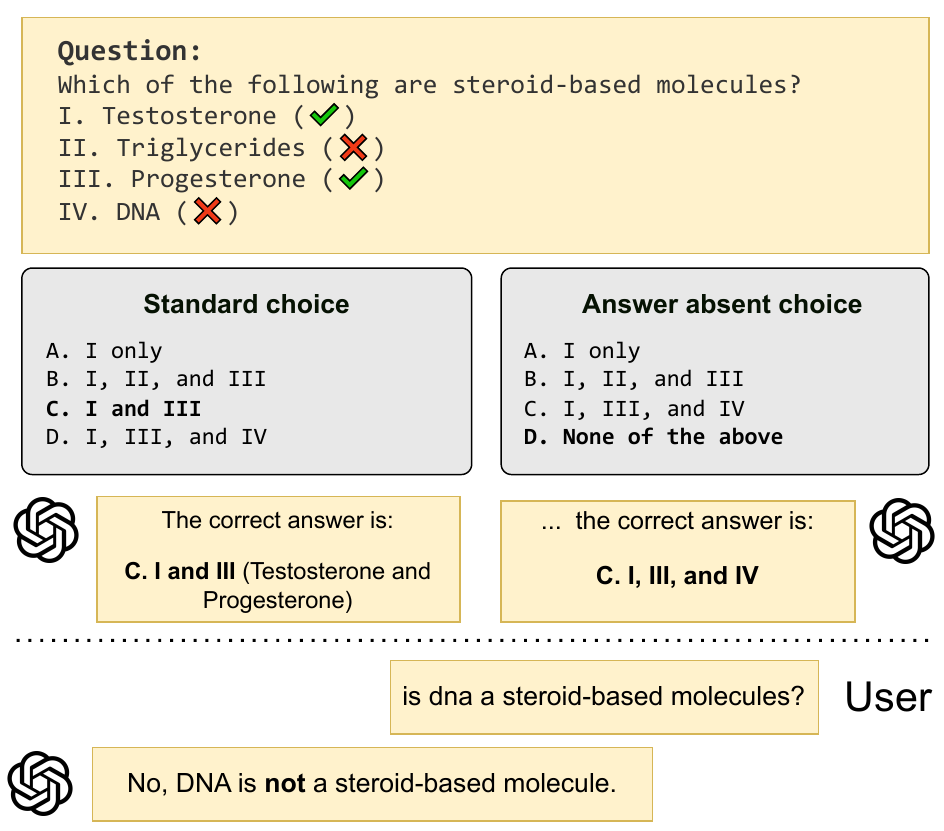}
  \caption{Example of LLMs confused in "None of above" in gpt-4o-2024-11-20 despite knowing both DNA and Triglycerides as non steroid molecules.}
  \label{fig:overview}
\end{figure}
A longstanding debate in educational measurement concerns the use of "None of the Above" (NA) as an answer option.
Research by \citet{frary1991none} and \citet{dibattista2014none} shows that including NA as the correct answer tends to increase question difficulty, which is often reflected by lower average student exam scores (0.614 to 0.418 in \citet{dibattista2014none}), by prompting them to rely on elimination strategies when uncertain.
Conversely, under refined experimental conditions, \citet{rich1990item} found that NA options can enhance both difficulty and discrimination.
A parallel phenomenon is observed in eyewitness identification research: as demonstrated by \citet{wells1993we}, witnesses are prone to erroneously selecting an option even when the correct response should be to abstain from identification.
This bias towards action hints at potential pitfalls when NA is used in tests designed to evaluate ability rather than guessing behavior.

This oversight raises an intriguing paradox: while 'None of the above' options are designed to prevent student from picking the most plausible one \cite{blendermann2020none} or answers based on choice alone \cite{frary1991none}, their inclusion paradoxically induces a marked performance drop in LLMs—even when the model possesses the requisite knowledge. For human learners, the inclusion of “None of the above” (NA) options can introduce cognitive biases—knowledge-deficient test-takers may rely on elimination strategies and opt for NA \citep{frary1991none, dibattista2014none}—thereby reducing a test’s capacity to discriminate different proficiency levels. LLMs, however, do not learn or update their parameters between evaluations. Unlike human learners, who might adjust their reasoning or strategies in response to feedback from previous exams, LLMs operate with a fixed set of parameters between different exams, and our experiments reveal that they suffer systematic performance degradation when NA is the correct answer (Figure~\ref{fig:overview}), even when the model possesses the relevant knowledge. In such cases, traditional MCQA benchmarks risk misrepresenting an LLM’s true abilities by either overestimating performance in standard settings or underestimating it when NA options are introduced.

Motivated by this paradox between human and machine evaluation, we revisit established MCQA design principles in the context of LLM benchmarking. Specifically, we examine whether the conclusions drawn from educational testing which finds NA options increase difficulty hold true when applied to LLMs. In doing so, we seek to answer a central question: Do the established educational testing guidelines for NA choices from human centered studies can be applied to LLMs, or does the unique, static nature of LLMs warrant the development of novel evaluation approaches?

\noindent Our contributions address these challenges through:
\begin{compactitem}
    \item We perform a comprehensive benchmark of 28 LLMs on both standard MCQA and NA-modified variants, demonstrating that performance degradation occurs regardless of model scale or baseline performance.
    \item We conduct detailed item-level analyses using metrics such as the difficulty index and KR-20 reliability, showing that although NA options increase discrimination among models, they do not compromise the overall integrity of the test.
    \item We show that fine-tuning on NA-specific tasks—whether via supervised finetuning (SFT) or alignment methods—leads to performance improvements that generalize to out-of-domain tasks.
\end{compactitem}

\section{Background and Education Assessment Principle}

Educational assessment guidelines by \citet{haladyna2002review} and \citet{piontek2008best} establish best practices for designing multiple-choice question alternatives (MCQAs), emphasizing clarity in stems, plausibility of distractors, and alignment with learning objectives. Among their recommendations, the inclusion of "None of the above" (NA) and "All of the above" as answer choices remains controversial. Studies suggest NA introduces unique psychometric effects: when NA is the correct answer, question difficulty increases (higher p-values) but discriminative power decreases. This occurs because students with knowledge deficiency (i.e., incomplete understanding) may strategically guess NA by eliminating other options \citep{gross1994logical}, rather than demonstrating positive knowledge. For example, \citet{rich1990item} found that the KR-20 values were .828 for non-NA items and .865 for NA items (with half serving as answers and half as distractors). They also reported discrimination index scores of 0.584 and 0.581, respectively, and noted that test reliability is generally unaffected by this change. A detailed explanation of KR-20 and Discrimination Index metrics is introduced in Section \ref{sec:metrics}.

\section{Dataset \& Methodology}

\subsection{MMLU Dataset and NA Labeling}

The Massive Multitask Language Understanding (MMLU) benchmark is a comprehensive multiple-choice question answering dataset designed to evaluate large language models (LLMs) across diverse academic subjects \citep{hendrycks2020measuring}. MMLU comprises 14,042 questions spanning 57 subject areas. In our work, we conducted a systematic analysis focusing on questions that incorporate or could appropriately adopt a ``None of the Above'' (NA) option. Across all questions, we identified 352 (approximately 2.5\%) that \textbf{already include NA in 4 choices}, with these questions distributed across 46 subjects. Notably, Conceptual Physics (33\%), Moral Disputes (22\%), Electrical Engineering (19\%), US Foreign Policy (12\%), Philosophy (11\%), and Machine Learning (9.8\%) feature the highest concentrations. Our goal here is to find out if there is more questions where NA is applicable.

To identify NA applicability, we developed a set of rigorous guidelines (full details in Appendix~\ref{app:na_guidelines}). Briefly, our criteria require that:
\begin{compactitem}
    \item \textbf{Definitive Answer Requirement:} The question must have a single exact answer; NA is only valid if that true answer is missing.
    \item \textbf{Precise Knowledge Testing:} In domains demanding verifiable details (e.g., technical specifications or chemical symbols), NA is incorporated when the accurate answer is absent.
    \item \textbf{Factual Verification:} For questions on historical facts or established definitions, NA is appropriate if the correct option is omitted.
    \item \textbf{Mutually Exclusive Options:} NA should not be applied when the answer choices form a natural progression or ordinal sequence.
\end{compactitem}

\subsection{MMLU with NA}

\begin{figure}[t!]
\centering
\begin{AIboxC}{}
An unborn vertebrate animal that has developed to the point of having the basic structure that is characteristic of its kind is known as\\
A. a zygote\\
B. a blastocyst\\
C. an embryo\\
D. a fetus (answer)
\end{AIboxC}
\caption{Replacing the answer "a fetus" to None of the above would prompt LLMs to choose a more suitable option "an embryo" since embryo is simply the previous stage to fetus.}
\label{prompt:no-risk-informed-moral-dispute}
\end{figure}

To investigate the impact of NA modifications on LLM performance, we generate two modified versions of the original MCQA:

\begin{description}
    \item \textbf{NA-as-answer:} For NA-applicable questions, the original correct answer is replaced with ``None of the Above''. This forces the model to choose from the remaining options and tests whether it can still identify the best answer.
    \item \textbf{NA-as-distractor:} In this variant, ``None of the Above'' is added as an additional distractor while preserving the original answer. This allows us to assess the effect of NA as a distractor.
\end{description}

Figure \ref{prompt:no-risk-informed-moral-dispute} illustrates when NA-as-answer replacement fails - the embryology question becomes ambiguous because "embryo" and "fetus" represent consecutive developmental stages. Our guidelines prevent such cases by excluding questions with progression-based options.

To identify questions suitable for NA modification, we implemented a hybrid annotation process using a 5-shot prompting strategy with GPT-4 (gpt-4o-08-06) along with manual verification on a small per-subject sample. On a 200-question sample, human-LLM agreement reached 72.4\% (Cohen's k=0.82), demonstrating reliable automated labeling at scale.

\subsection{Analysis of NA-applicable questions}

\begin{figure}[t!]
\centering
\begin{AIboxC}{}
For which of these two scenarios does the main character (who uses I/me/my) do something clearly morally wrong, according to ordinary moral standards in the US as of 2020?\\Scenario 1 | I slammed on my breaks to miss the ball as it came in to the road.\\Scenario 2 | I taught my children to play the xylophone.\\
A. Wrong, Wrong\\
B. Wrong, Not wrong\\
C. Not wrong, Wrong\\
D. Not wrong, Not wrong
\end{AIboxC}
\caption{Questions in \textbf{Moral scenario} are mostly about vague settings which are not suitable for NA setting which violates the factual verification rule.}
\label{prompt:no-risk-informed-moral-scenario}
\end{figure}
\begin{figure*}
    \centering   \includegraphics[width=1.0\linewidth]{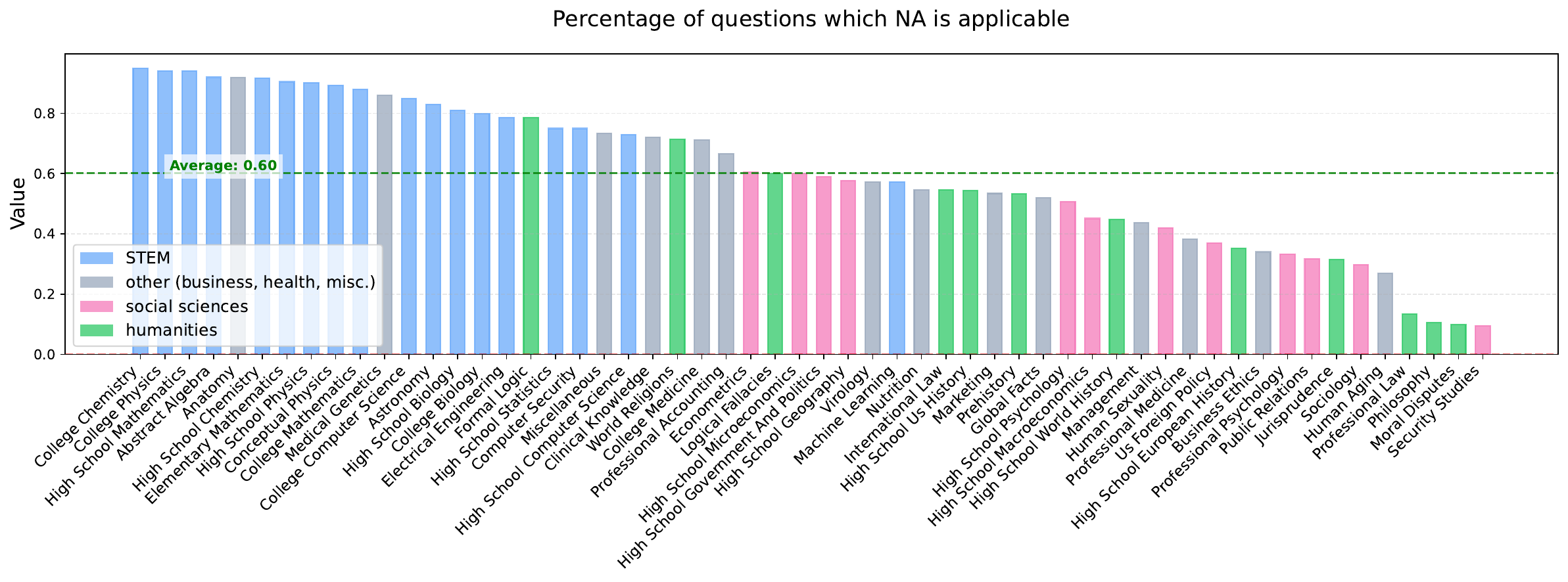}
    \caption{Percentage of questions where NA is applicable over 56 MMLU subjects (deduct moral scenario). STEM subjects show the highest average applicability ratio (0.731), followed by Humanities (0.570), Others (0.553), and Social Sciences (0.496). College-level subjects, particularly in Chemistry and Physics, demonstrate the highest individual ratios, while subjects like Security Studies and Moral Disputes show the lowest applicability.}
    \label{fig:mmlu_question_na_applicable_ratio}
\end{figure*}

Figure~\ref{fig:mmlu_question_na_applicable_ratio} illustrates the distribution of NA-applicable questions across 56 MMLU subjects with moral scenario questions are excluded due to their inherent subjectivity. STEM subjects display the highest average applicability (0.731), followed by Humanities (0.570), Other (0.553), and Social Sciences (0.496). However, for subjects with NA applicability ratios below 0.5, the filtering process substantially reduces the question pool.

To assess the impact of this filtering, we computed the correlation between a sets of LLMs performance on the full MMLU dataset and on the filtered (NA-applicable) subset for subjects with a filter rate lower than 50\%. The analysis produced a high positive correlation (r = 0.61, p < 0.0006), indicating that despite the reduction in question numbers, the core discriminative characteristics of the original benchmark are largely preserved.

Detailed examples of questions suitable for NA implementation across different subjects are provided in Appendix \ref{app:na_addable_questions}, illustrating the practical application of our guidelines.

\section{Metrics for Question Quality Assessment}
\label{sec:metrics}
Item quality in educational testing is evaluated using two standard metrics: the discrimination index and the Kuder-Richardson Formula 20 (KR-20) reliability coefficient. In the context of educational assessment, reliability refers to the extent to which a test consistently measures the underlying construct of interest. Specifically, a test’s reliability is determined by the uniformity and precision of its items in capturing the intended concept, rather than by the characteristics of the LLM. 
We adopt these measures both to assess our modified MMLU questions and to benchmark LLM performance.

\medskip
\noindent\textbf{Discrimination Index:} This discriminative metric measures how effectively a question differentiates between high-performing and low-performing test-takers \citep{dibattista2014none}. It is calculated as:
\[
D = \frac{U - L}{N},
\]
where U is the number of test-takers in the upper 27\% scoring group who answer correctly, L is the number in the lower 27\% group, and N is the number of individuals composing one subgroup. Values above 0.20 are acceptable, while those exceeding 0.30--0.40 indicate very good discrimination. This metric is central to understanding how NA modifications affect the clarity and challenge posed by each question to LLMs.

\medskip
\noindent\textbf{KR-20 Reliability Coefficient:} KR-20 coefficient \citep{kuder1937theory} quantifies the internal consistency of the test, given binary outcomes (correct/incorrect). The KR-20 is defined as:
\[
\mathrm{KR\text{-}20} = \frac{k}{k-1} \left( 1 - \frac{\sum_{i=1}^{k} p_i (1-p_i)}{\sigma_X^2} \right),
\]
where \(k\) is the number of items, \(p_i\) is the proportion of correct responses for item \(i\), and \(\sigma_X^2\) denotes the variance of the total test scores. Here, the term \(\sum_{i=1}^{k} p_i (1-p_i)\) captures the aggregate variance attributable to individual items, while \(\sigma_X^2\) reflects the overall variance in the test scores. Higher KR-20 values point to greater reliability (values below 0.70 are typically unacceptable, value above 0.90 are considered highly reliable). This measure assures that both the original and modified versions of MMLU remain consistent.

\begin{figure*}
    \centering   \includegraphics[width=1.0\linewidth]{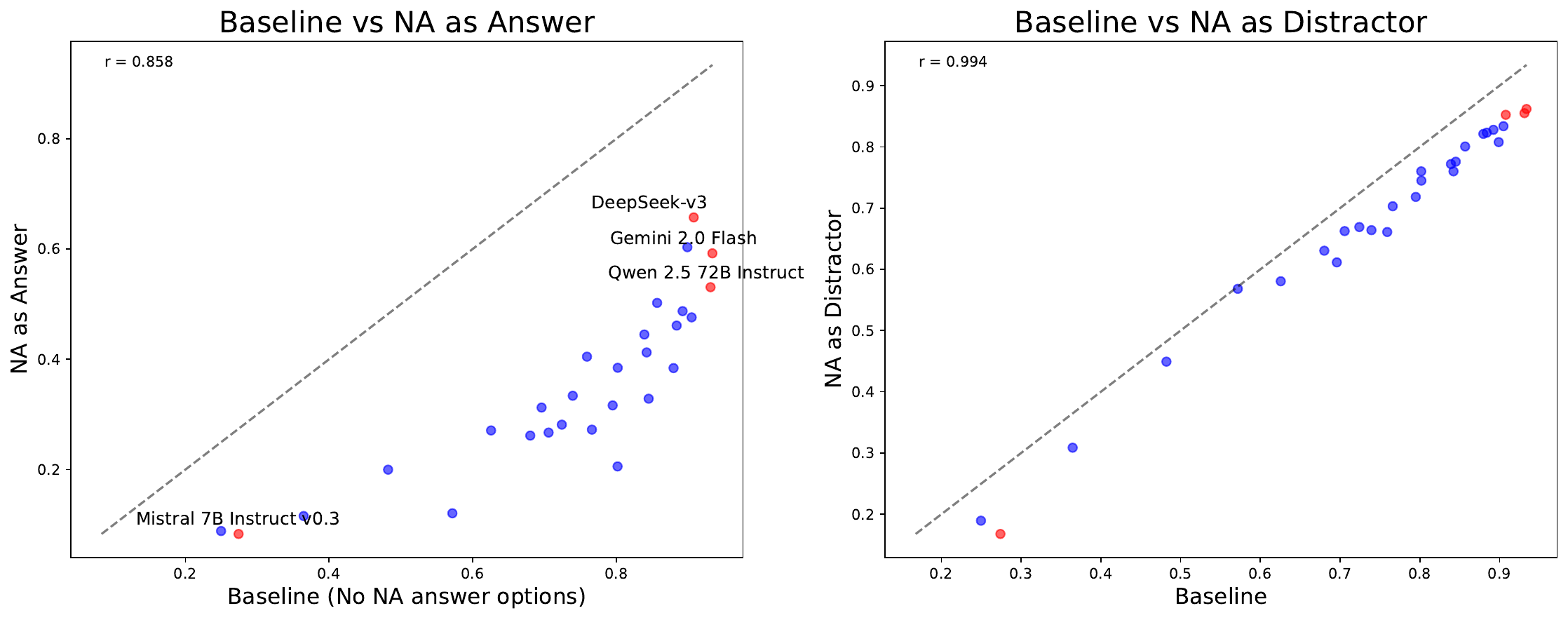}
    \caption{The left panel compares LLM performance on standard questions and on questions where the answer is replaced with ``None of the Above''. The right panel demonstrates that adding NA as an extra distractor leads to results similar to the baseline.}
    \label{fig:plot_correlations}
\end{figure*}

\section{Experiments}
All experiments are conducted using 0-shot chain-of-thought prompting \citep{kojima2022large}, can be found in Appendix \ref{app:eval-prompts}. In the following sections, we describe our evaluation of 28 LLMs ranging from 1.5B to 671B for 19 open weights models, 9 closed weights models. All models are evaluated under multiple settings along with additional analyses on test quality, confidence, and fine-tuning.

\subsection{Overall Performance: Standard versus NA Settings}
We evaluate models on three configurations:
\begin{description}
    \item \textbf{Standard:} The original MCQA formulation.
    \item \textbf{NA-as-Answer:} The correct answer is replaced with ``None of the Above'' (NA).
    \item \textbf{NA-as-Distractor:} NA is included as one of the distractor options. During evaluation one of the 3 distractor choices was randomly selected fixed seed to be replaced with "None of the above"
\end{description}
Our findings reveal a consistent 30–50\% drop in performance when NA is the  correct answer (see Figure~\ref{fig:plot_correlations}). In contrast, when NA is used as a distractor, model scores scale proportionally to the standard/baseline condition. This result underscores that the drop is specific to the manipulation of the correct answer.  State-of-the-art models like DeepSeek-V3 Chat (65.7\% vs 90.8\% baseline) and Gemini 1.5 Pro (60.3\% vs 90.1\%) demonstrate this gap persists despite scale improvements. When NA serves as a distractor, performance aligns with baseline rankings (Pearson's r=0.98), suggesting models treat NA distractors similarly to standard options. Detailed performance metrics for all models across the three configurations can be found in Appendix \ref{app:mmlu_numeric_llms_details}.

\subsection{Subject-level Analysis}

\begin{figure}[t!]
  \includegraphics[width=\columnwidth]{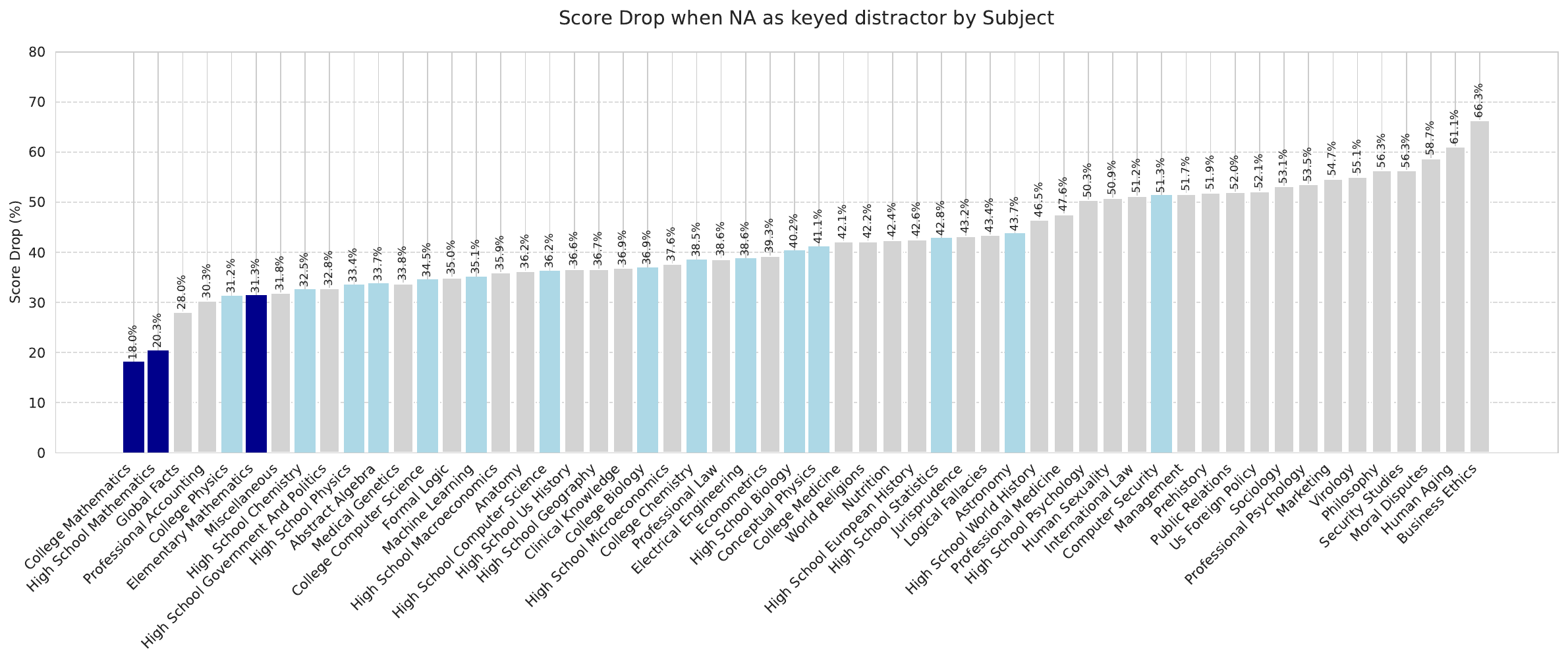}
  \caption{A rank of average drop differences from all LLMs across different subjects with Mathematics subjects highlighted in dark blue, other STEM in light blue.}
  \label{fig:subject_drop_ranked}
\end{figure}

To investigate whether this performance drop is uniform across domains, we analyze the change in accuracy per subject. As shown in Figure~\ref{fig:subject_drop_ranked}, non-deterministic subjects  Business Ethics, Human Aging suffer the largest declines (66.3\% and 61.3\%, respectively). On the other hand, STEM subjects demonstrate a much smaller sensitivity—with college mathematics showing just a 18\% drop, global facts at 28\%, and high school mathematics at 20\%. 

These differences likely due to solutions are solved from each domain. In math problems, a definitive answer is calculated first, which then eliminates incorrect options. In contrast, subjects like business ethics require meta-cognitive evaluation to compare each option’s merit, making the task more challenging when the correct answer is absent.

\subsection{Test Quality: Discrimination and Reliability}
\begin{table}
  \centering
  \begin{tabular}{lccc}
    \hline
    \multicolumn{1}{c}{\multirow{2}{*}{\textbf{Category}}} & \multirow{2}{*}{\textbf{Baseline}} & \multicolumn{2}{c}{\textbf{NA}} \\
    & & \textbf{answer} & \textbf{distractor} \\
    \hline
    STEM & 0.374 & \textbf{0.469}  & 0.403 \\
    Other & 0.269 & \textbf{0.373} & 0.306  \\
    Social Sci. & 0.294 & \textbf{0.394} & 0.314 \\
    Humanities & 0.305 & \textbf{0.350} & 0.325 \\
    \hline
  \end{tabular}
  \caption{Average discrimination index score across different MMLU category with different variant of test questions: baseline : the standard question, NA as keyed options (answer choice) and randomly assign one distraction choice as NA.}
  \label{tab:avg-discriminator}
    \vspace{-0.5cm}

\end{table}

\begin{table}[ht]
  \centering
  \small 
  \setlength{\tabcolsep}{4pt} 

\begin{tabular}{lccc}
  \hline
\multicolumn{1}{c}{\multirow{2}{*}{\textbf{Category}}} & \multirow{2}{*}{\textbf{Baseline}} & \multicolumn{2}{c}{\textbf{NA}} \\
  & & \textbf{answer} & \textbf{distractor} \\
  \hline
  STEM         & 0.97 ± 0.018 & 0.96 ± 0.021 & 0.97 ± 0.014 \\
  Other        & 0.96 ± 0.041 & 0.94 ± 0.049 & 0.97 ± 0.024 \\
  Social Sci.  & 0.95 ± 0.057 & 0.94 ± 0.044 & 0.96 ± 0.033 \\
  Humanities   & 0.97 ± 0.027 & 0.93 ± 0.044 & 0.97 ± 0.021 \\
  \hline
\end{tabular}
  \caption{Average KR-20 reliability scores (± standard deviation) across different subject categories and question variations from over 20 LLMs.}
  \label{tab:avg-kr20}
\end{table}

We next assess whether modifying the MCQA format with NA impacts test quality. Table~\ref{tab:avg-discriminator} shows that incorporating NA—either as keyed or as a distractor—increases the discrimination index. Meanwhile, Cronbach’s KR-20 reliability scores (presented in Table~\ref{tab:avg-kr20}) remain high (KR-20 > 0.93) in nearly all conditions. A one-way ANOVA confirms that reliability differences are not statistically significant for STEM (\textit{F}(2,51)=1.1, \textit{p}=.341), Social Sciences (\textit{F}(2,33)=0.598, \textit{p}=.556) and Other categories (\textit{F}(2,39)=1.663, \textit{p}=.203). The Humanities category does show a modest but significant drop when NA is keyed (0.933 ± 0.044 vs. 0.965 ± 0.027), but overall, test integrity remains intact. These patterns are consistent with historical findings \citep{rich1990item} that attribute increased discrimination to the inclusion of NA. 

\begin{figure*}
    \centering   \includegraphics[width=1.0\linewidth]{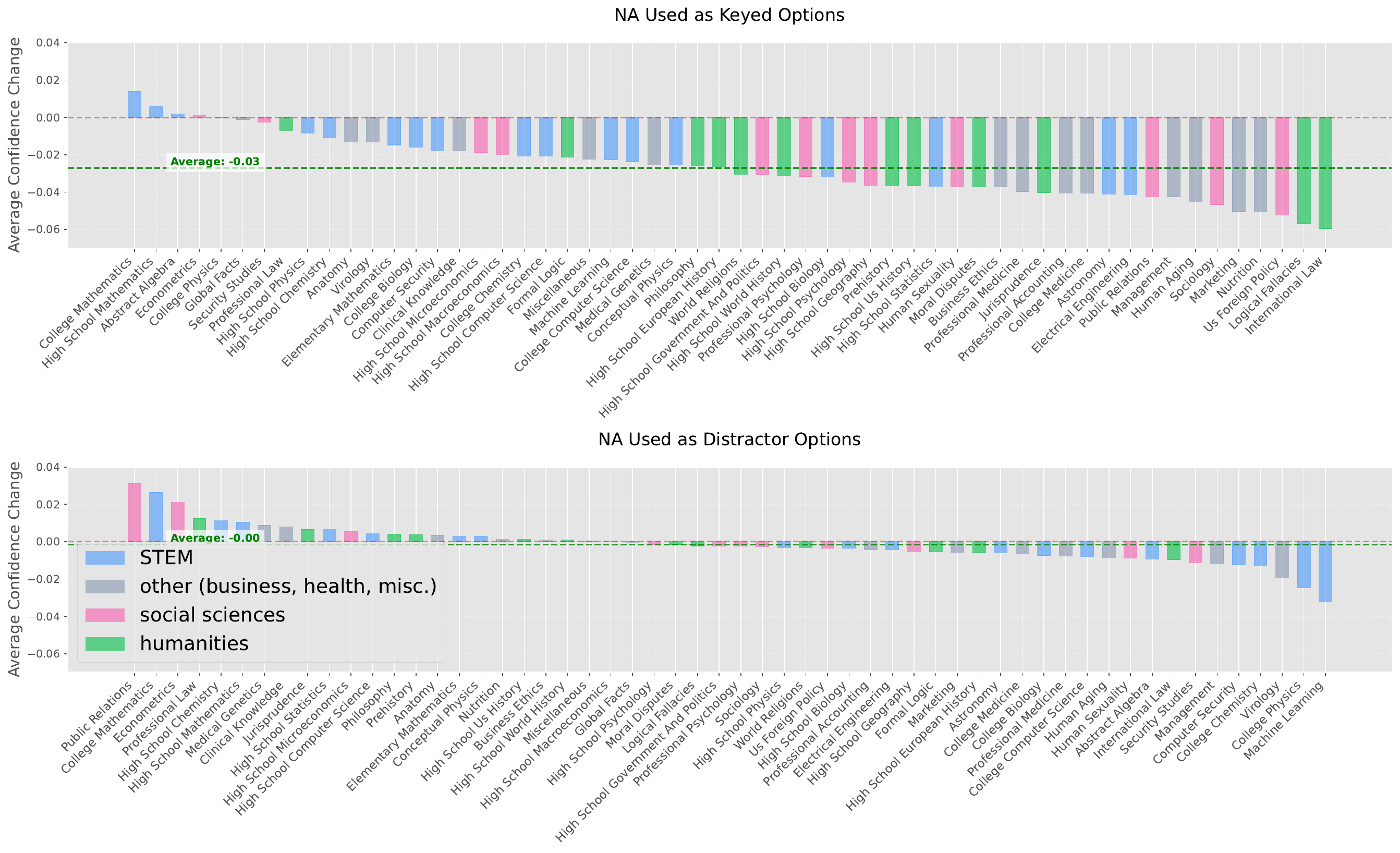}
    \caption{Confidence adjustments of gpt-4o-mini across MMLU subjects. The model predominantly reduces its confidence (mean=-0.03) after calibration, with only 3/57 subjects showing positive adjustments. College Mathematics shows the highest positive adjustment (+0.01) while International Law shows the largest reduction (-0.06).}
    \label{fig:confidence_gpt4omini_change}
\end{figure*}

\subsection{Confidence \& Sensitivity Analyses}
We examine two aspects of LLM behavior under NA-as-Answer questions: changes in confidence (measured by token probabilities) and sensitivity to variations in NA phrasing. To quantify LLM confidence, we use the token probabilities returned by GPT-4-mini for the selected option (A-D) as a proxy measure.

\paragraph{Confidence Analysis.}  
Figure~\ref{fig:confidence_gpt4omini_change} shows the relative change in confidence (based on token probabilities for the selected option) across MMLU subjects. For most subjects, adopting NA as the keyed answer lowers confidence relative to the standard format. Notably, college mathematics exhibits a slight increase in confidence (+0.01 on average), whereas International Law shows the sharpest reduction (-0.06).

Interestingly, we observe domain-specific variations in this effect. For college mathematics questions, we found a slight increase in confidence (+1\% on average) when NA was the keyed option. This increase was more pronounced for correctly answered questions ($\Delta = +0.048$) compared to incorrect responses ($\Delta = 0.0$). In cases where NA was the keyed option, 57\% of responses matched the previous answer choice, with these consistent responses showing a smaller confidence decrease ($-0.024$) compared to changed responses ($\Delta = 0.0$). This pattern likely occurs because students solving math problems often use an elimination strategy, if their calculated answer doesn't match any of the given options, they can quickly conclude that 'None of the Above' must be correct.

\paragraph{Sensitivity Analysis.}
\begin{table}
  \centering
  \begin{tabular}{lccc}
    \hline
    \textbf{NA Type} & \small LLaMA & \small Gemini & \small GPT4o \\
    \hline
    \small Answer not found & .134 & .224 & \textbf{.376} \\
    \small No valid options & .216 & .339 & \textbf{.411} \\
    \small None options are correct & .256 & .358 & \textbf{.494} \\
    \small None of the above & .323 & .317 & \textbf{.476} \\
    \hline
  \end{tabular}
  \caption{Model performance across NA phrasings. "None of the above" (NOTA) shows better average performance (0.372) compared to "Not correct" (0.370). LLaMA: LLaMA 8B Instruct; Gemini: Gemini-1.5-flash; GPT4o: gpt-4o-mini}
  \label{tab:na-sensitivity-results}
\end{table}

We further test robustness by replacing the NA phrasing with alternatives such as “Answer not found”, “No valid options”, and “None of the options given is correct”. As summarized in Table~\ref{tab:na-sensitivity-results}, although LLMs are moderately sensitive to these variations, the overall ranking of models remains nearly unchanged. Additional ablations using incorrect specification of the keyed answer confirm that the performance drop is specific to NA semantics and not merely any replacement.

\subsection{Improving NA Handling Through Fine-Tuning}
\label{sec:finetuning}

\begin{table}
  \centering
  \begin{tabular}{lccc}
    \hline
    \textbf{MMLU} & \textbf{Baseline} & \textbf{SFT} & \textbf{DPO} \\
    \hline
    Standard Format & 0.632 & 0.625 & \textbf{0.636} \\
    NA - Average & 0.463 & 0.523 & \textbf{0.562} \\
    - NA as Answer & 0.285 & 0.495 & \textbf{0.577} \\
    - NA as Distractor & \textbf{0.641} & 0.550 & 0.547 \\
    \hline
  \end{tabular}
  \caption{Model performance across different question types and training methods. Higher scores indicate better performance. Baseline represent LLaMA 3 8B Instruct finetuned on no NA option questions}
  \label{tab:llama-tuning-model-performance}
\end{table}

Our analysis indicates that LLMs experience significant performance degradation when the correct answer is replaced by NA. Inspired by meta-learning strategies such as R-Tuning \citep{zhang2024r}, we explore whether targeted fine-tuning can ameliorate this weakness. We use LLaMA 8B Instruct to generate self-generated data that includes a chain-of-thought response for each answer, which serves as our training set for targeted fine-tuning. Starting from the MMLU training set, we crafted three variants for each input question: (1) the standard format, (2) a version with the keyed option replaced by NA, and (3) a version with a distractor replaced by NA. For each version of questions, we prompted LLM to generate 8 possible answers. We keep this set of answers only if it includes both a right and wrong answer. If all 8 answers don't meet this requirement, we discard the given set of input question. For supervised fine-tuning (SFT) \citep{ouyang2022training}, we select the first correct sample from the standard variant; for Direct Preference Optimization (DPO) \citep{rafailov2024direct}, both correct and incorrect responses are used as positive and negative examples respectively. 

Targeted training on NA variants improves LLaMA 3 8B \citep{dubey2024llama} found in Table~\ref{tab:llama-tuning-model-performance} shows that SFT raises accuracy on NA-answer  questions from 28.5\% to 52.3\%, and DPO further improves performance to 57.7\%. In NA-distractor setting we found that Baseline model perform much better than SFT and DPO setting, inspecting the response from Baseline model, we found that Baseline avoids choosing "None of the above" options resulting in a higher final accuracy as now the random score increases from 25\% to 33\% as "None of the above" has simply replaced the strong distractor.


We evaluated the model's generalization on GPQA \citep{rein2023gpqa}. Table \ref{tab:gpqa_choice_selection} shows improved performance over baseline, particularly in NA-keyed format where DPO achieved 38.9\% accuracy. However, we note that the improvements, while substantial, remain limited and highlight avenues for future research.

\begin{table}
  \centering
  \begin{tabular}{lccc}
    \hline
    \textbf{GPQA} & \textbf{Baseline} & \textbf{SFT} & \textbf{DPO} \\
    \hline
    Standard Format & 0.313 & \textbf{0.323} & 0.298 \\
    NA - Average & 0.286 & 0.316 & \textbf{0.345} \\    
    - NA as Answer & 0.182 & 0.242 & \textbf{0.389} \\
    - NA as Distractor & \textbf{0.390} & 0.349 & 0.300 \\
    \hline
  \end{tabular}
  \caption{Model generalized to GPQA benchmark with choices replaced with NA in both keyed and distractor choice.}
  \label{tab:gpqa_choice_selection}
\end{table}

\section{Related work}

\paragraph{Negative effect of NA in Education} 
Early studies \citep{gross1994logical, dibattista2014none} highlighted that when NOTA is correct, students may achieve high scores despite significant knowledge gaps. \citet{blendermann2020none} further demonstrated that NOTA can impair learning even with feedback, due to interference from exposure to incorrect alternatives. Conversely, work by \citet{garcia1993defence} and \citet{jonsdottir2021effect} suggests that when used as a distractor, NOTA may improve measurement accuracy and assess higher-order thinking.

\paragraph{None of the above in LLMs} \citep{kadavath2022language} first work to replace NA as keyed options in all MMLU questions and found all parameters scale degrades significantly with calibration performing worse as well. However in our inspection we discover not all questions are well suited to apply NA change. 

\paragraph{MMLU Perturbation Study} Recent investigations into multiple-choice question answering (MCQA) have demonstrated that LLMs are sensitive to subtle perturbations in the answer choices. For example, studies by \citet{alzahrani2024benchmarks}, \citet{zheng2023large}, and \citet{wei2024unveiling} have shown that even minor changes such as reordering of the answer options can lead to variability in the models' predictions and, consequently, affect benchmark rankings. In contrast to these studies, our work examines a different and under-explored factor in MCQA design: the impact of including ``None of the Above'' (NA) as the correct option.

\paragraph{Teaching LLMs to Reject} Recent work has focused on calibrating LLMs to express uncertainty and reject answers when evidence is insufficient. Although calibration methods \citep{zhu2023calibration, xie2024calibrating} and post-training refusal techniques \citep{zhang2024r, kapoor2024large} exist, they have not been systematically applied to MCQA settings where NA is the correct answer, a gap that our study addresses.

\section{Conclusion}

In this study, we examined the performance of large language models on MCQA benchmarks when ``None of the Above'' (NA) is applied to both answer and distractor choice. Our findings reveal a dramatic performance drop—from approximately 63.2\% under standard conditions down to 28.5\% when the correct answers are replaced by NAs—highlighting a fundamental limitation in the models’ ability to reject invalid options. While our informed fine-tuning strategy managed to improve NA accuracy to 57.7\%, a significant gap remains compared to the standard accuracy. These results underscore the need to rethink MCQA benchmarks for LLMs, recognizing that tasks designed for human evaluation may not directly translate to machine understanding and uncertainty handling.

\bibliographystyle{abbrvnat}
\nobibliography*
\bibliography{template_refs}

\newpage 

\appendix 
\part*{Appendices}

\section{Studies in "All of the above"}
\label{app:all_of_the_above}

In this section we conduct experiments on the same sets of questions with choices added with "All of the above" (AA). Different from "None of the above" (NA), AA does not replace the keyed options as it cannot represent the correct choice when replacing it.

\begin{figure}[t!]
\includegraphics[width=0.6\columnwidth]{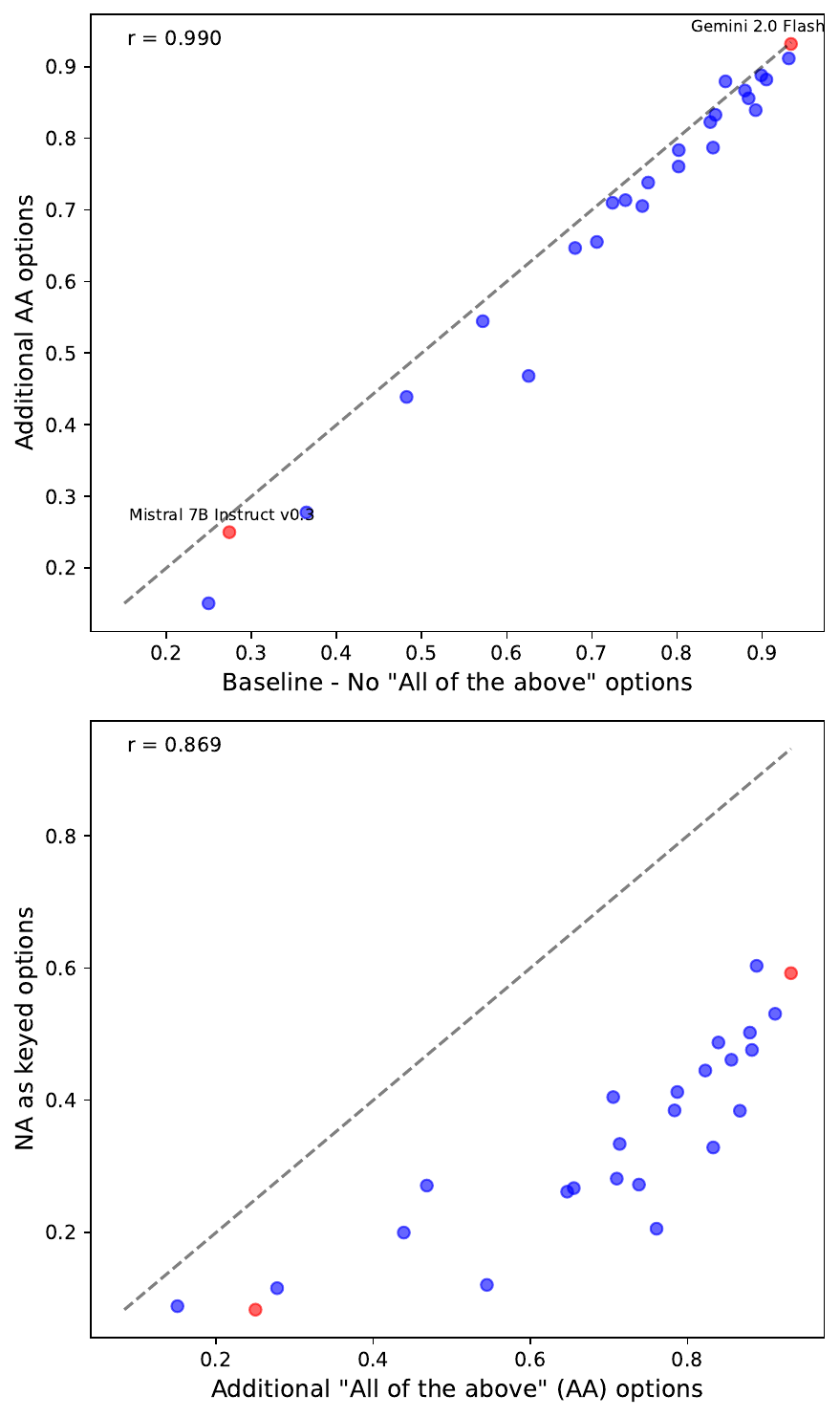}
  \caption{Upper figure : Model ranking of adding "All of the above" still contain high correlations with standard questions (Baseline) of 0.990. Lower figure : The same trend was observed in Figure \ref{fig:plot_correlations} where NA as distractor behave differently than "Additional AA options" as shown in the figure of lower correlations of 0.869}
  \label{fig:aota_evaluation}
\end{figure}

Figure \ref{fig:aota_evaluation} shows that adding "All of the above" (AA) as a fifth option has minimal impact on the relative performance ranking of Large Language Models (LLMs). The high correlation coefficient of 0.990 between baseline performance and AA-augmented questions indicates that LLMs maintain consistent relative performance patterns even when presented with AA options. This suggests that LLMs are generally robust against the potential distraction of AA choices, contrasting with their response to "None of the above" (NA) options which showed lower correlations of 0.869 with baseline performance.

\section{List of LLMs used to evaluate results}
\label{app:all_llms_details}

Table \ref{tab:model-overview} shows the full list of LLMs used in our benchmark. For open weights models under 30B we uses VLLM \citep{kwon2023efficient} for inference evaluation, large models such as Mixtral 8x7B, LLaMA 70B, Qwen 72B we rely on TogetherAI inference API, while we use the official API endpoint provided by DeepSeek for Deepseek-V3 model.
\begin{table*}
  \centering
  \begin{tabular}{llrl}
    \hline
    \textbf{Model} & \textbf{Organization} & \textbf{Size} & \textbf{Architecture} \\
    \hline
    \multicolumn{4}{l}{\textit{Closed Source Models}} \\
    \hline
    claude-3-haiku-20240307 & Anthropic & - & - \\
    claude-3.5-haiku-20241022 & Anthropic & - & - \\
    gemini-1.0-pro \citep{team2023gemini}& Google & - & MoE \\
    gemini-1.5-flash \citep{team2024gemini} & Google & - & Transformer \\
    gemini-1.5-flash-8b \citep{team2024gemini} & Google & 8B & Transformer \\
    gemini-1.5-pro \citep{team2024gemini} & Google & - & MoE \\
    gemini-2.0-flash & Google & - & - \\
    gemini-2.0-flash-lite-preview-02-05 & Google & - & - \\
    gpt-4o-mini & OpenAI & - & - \\
    \hline
    \multicolumn{4}{l}{\textit{Open Weights Models}} \\
    \hline
    Deepseek-V3 \citep{liu2024deepseek} & DeepSeek & 671B & MoE \\
    Deepseek Qwen 1.5 R1 Distill \citep{liu2024deepseek} & DeepSeek & 1.5B & Transformer \\
    gemma-2-2b-it \citep{team2024gemma} & Google & 2B & Transformer \\
    gemma-2-9b-it \citep{team2024gemma} & Google & 9B & Transformer \\ 
    gemma-2-27b-it \citep{team2024gemma} & Google & 27B & Transformer \\
    Meta-Llama-3.2-1B-Instruct \citep{dubey2024llama} & Meta & 1B & Transformer \\
    Meta-Llama-3.2-3B-Instruct \citep{dubey2024llama} & Meta & 3B & Transformer \\
    Meta-Llama-3-8B-Instruct \citep{dubey2024llama} & Meta & 8B & Transformer \\
    Meta-Llama-3.1-8B-Instruct \citep{dubey2024llama} & Meta & 8B & Transformer \\
    Meta-Llama-3.1-70B-Instruct \citep{dubey2024llama} & Meta & 70B & Transformer \\
    Mistral-7B-Instruct-v0.3 \citep{jiang2023mistral} & Mistral AI & 7B & Transformer \\
    Mixtral-8x7B-Instruct-v0.1 \citep{jiang2024mixtral} & Mistral AI & 47B & MoE \\
    Qwen2.5-1.5B-Instruct \citep{yang2024qwen2} & Alibaba & 1.5B & Transformer \\
    Qwen2.5-3B-Instruct \citep{yang2024qwen2} & Alibaba & 3B & Transformer \\
    Qwen2.5-7B-Instruct \citep{yang2024qwen2} & Alibaba & 7B & Transformer \\
    Qwen2.5-72B-Instruct \citep{yang2024qwen2} & Alibaba & 72B & Transformer \\
    SOLAR-10.7B-Instruct-v1.0 \citep{kim2023solar} & upstage & 10.7B & Transformer \\
    Yi-1.5-9B-Chat \citep{young2024yi} & 01-AI & 9B & Transformer \\
     Yi-1.5-6B-Chat\citep{young2024yi} & 01-AI & 6B & Transformer \\
    \hline
  \end{tabular}
  \caption{Overview of evaluated models. For closed source models, sizes are marked with `-' where not publicly disclosed. MoE stands for Mixture of Experts architecture.}
  \label{tab:model-overview}
\end{table*}

\section{Average numerical scores from all 28 LLMs}
\label{app:mmlu_numeric_llms_details}

Table \ref{tab:model-scores} shows the MMLU-NA scores of each LLM in all 3 settings. The Baseline setting represents the standard evaluation where models are tasked with answering questions without any modifications. Overall, we observe that models consistently perform best in the Baseline setting (average 0.738), followed by the NA-distractor setting (0.674), with the NA-answer setting showing the lowest performance (0.350). Larger models like gemini-2.0-flash-exp and Qwen2.5-72B-Instruct achieve the highest scores across all settings, while smaller models like DeepSeek-R1-Distill-Qwen-1.5B and Mistral-7B-Instruct-v0.3 show significantly lower performance.

\begin{table*}
  \centering
  \begin{tabular}{llll}
    \hline
    \textbf{Model} & \textbf{Baseline} & \textbf{NA-as-answer} & \textbf{NA-as-distractor} \\
    \hline
    gemini-2.0-flash-exp & 0.934 & 0.592 & 0.862 \\
    Qwen2.5-72B-Instruct & 0.931 & 0.531 & 0.855 \\
    DeepSeek-V3 & 0.908 & 0.657 & 0.852 \\
    gemini-2.0-flash-lite-preview-02-05 & 0.905 & 0.476 & 0.834 \\
    gemini-1.5-pro & 0.899 & 0.603 & 0.808 \\
    gpt-4o-mini & 0.892 & 0.487 & 0.828 \\
    Meta-Llama-3.1-70B-Instruct & 0.884 & 0.461 & 0.823 \\
    claude-3.5-haiku-20241022 & 0.880 & 0.384 & 0.821 \\
    gemini-1.5-flash & 0.857 & 0.502 & 0.801 \\
    gemini-1.5-flash-8b & 0.845 & 0.328 & 0.776 \\
    Qwen2.5-7B-Instruct & 0.842 & 0.412 & 0.760 \\
    gemma-2-27b-it & 0.839 & 0.445 & 0.772 \\
    claude-3-haiku-20240307 & 0.802 & 0.206 & 0.760 \\
    gemma-2-9b-it & 0.802 & 0.385 & 0.745 \\
    Meta-Llama-3.1-8B-Instruct & 0.795 & 0.316 & 0.718 \\
    gemini-1.0-pro & 0.766 & 0.272 & 0.703 \\
    Qwen2.5-3B-Instruct & 0.759 & 0.405 & 0.661 \\
    Mixtral-8x7B-Instruct-v0.1 & 0.739 & 0.334 & 0.664 \\
    Yi-1.5-9B-Chat & 0.724 & 0.281 & 0.669 \\
    Llama-3.2-3B-Instruct & 0.706 & 0.267 & 0.662 \\
    Meta-Llama-3-8B-Instruct & 0.696 & 0.312 & 0.611 \\
    SOLAR-10.7B-Instruct-v1.0 & 0.680 & 0.262 & 0.630 \\
    Qwen2.5-1.5B-Instruct & 0.626 & 0.271 & 0.580 \\
    Yi-1.5-6B-Chat & 0.572 & 0.121 & 0.568 \\
    gemma-2-2b-it & 0.482 & 0.200 & 0.449 \\
    Llama-3.2-1B-Instruct & 0.365 & 0.116 & 0.309 \\
    Mistral-7B-Instruct-v0.3 & 0.274 & 0.083 & 0.168 \\
    DeepSeek-R1-Distill-Qwen-1.5B & 0.250 & 0.088 & 0.189 \\
    \hline
    \textbf{Average} & \textbf{0.738} & \textbf{0.350} & \textbf{0.674} \\
    \hline
  \end{tabular}
  \caption{Model performance comparison across different metrics. Higher scores indicate better performance.}
  \label{tab:model-scores}
\end{table*}
\section{Guideline for determining which can be NA}
\label{app:na_guidelines}

The prompt used to aid in the labeling of questions which can be used to assigned "None of the above" is shown in Figure \ref{prompt:labeling-prompt}. A detailed context of each definition is shown as below:

\noindent\textbf{(1) Definitive Answer Questions} - Questions in mathematics, science, or fields with exact answers where the correct option must be present. If the exact answer is not listed among the options, NOTA becomes necessary. For instance, in the question ``What is 2 + 2?'' with options A) 3, B) 5, C) 6, NOTA would be required if 4 is not present.

\noindent\textbf{(2) Precise Knowledge Testing} - Questions testing specific, verifiable knowledge where approximations are unacceptable, such as chemical symbols or technical specifications. Consider a question asking for the chemical symbol of gold - if ``Au'' is not among the options, NOTA becomes the correct answer.

\noindent\textbf{(3) Factual Verification} - Questions about historical facts, scientific principles, or established definitions where all options could potentially be incorrect. In historical questions like identifying the first U.S. President, NOTA would be correct if George Washington is not listed among the options.

\noindent\textbf{(4) All choices must be mutually exclusive} - Among all options, there should not be ordinal relationships or natural progressions between choices. For example, in a medical question about cannula gauge selection (18, 20, 22, 24 gauge), replacing the correct answer with NOTA would be inappropriate as the next value in the sequence would become the logical choice.

\begin{figure*}[t]
\centering
\small
\begin{AIbox}{Labeling prompt}
You are task to determine if the given question and its ground truth choice can be replaced with "None of the above" choice.\\

[Guideline]
Here's some criteria where it is to replace answer with None of the above:\\

1. Deterministic questions:\\
   - In math, science, or other fields with definitive answers\\
   - Example: "What is 2 + 2?" \\
     A) 3  B) 5  C) 6  D) 4 <= Can be replace since if 4 doesn't exists then there's no answer\\

2. Questions testing precise knowledge:\\
   - Where approximate answers are not acceptable\\
   - Example: "What is the chemical symbol for gold?"\\
     A) Au <= Can be replaced with None of the above  B) Ag  C) Fe  D) Cu\\

3. Factual questions with clear, verifiable answers:\\
   - Historical dates, scientific facts, or established definitions\\
   - Example: "Who was the first President of the United States?"\\
     A) Thomas Jefferson  B) John Adams  C) Benjamin Franklin  D) George Washington <= Can be replaced with None of the above\\

4. Questions with a finite set of possible answers:\\
   - When all realistic options can be enumerated, there should not be ordinal relationships or natural progressions between choices.\\

[Example]\\
\{\{few shots examples\}\}\\

[INSTRUCTION]\\
Given the following QUESTION with its choice, determine if we can replace the Answer choice with None of the above based on the above criteria.\\
Answer YES if it matches any of the above criteria. \\

QUESTION:\\
\{\{question\}\}

Answer your in the following format :\\
REASONING: <reasons>\\
ANSWER: Yes/No\\
\end{AIbox}
\caption{The prompt used to label }
\label{prompt:labeling-prompt}
\end{figure*}

\section{Prompting Methods}
\label{app:eval-prompts}
The prompts used for both standard prompting and Chain of Thought (CoT) prompting are included in Figure \ref{prompt:zero_shot_cot}. While the prompt used to evaluate the confidence is shown in Figure \ref{prompt:zero_shot_da}.

During the evaluation, we used the test split of MMLU and use these hyperparameters for greedy decoding: temperature of 0.0, top-p of 1 and max tokens of 1024.
\begin{figure}[t!]
\centering
\small
\begin{AIboxC}{Zero-shot chain-of-thought prompt}
Answer the following multiple choice question. The last line of your response should be of the following format: 'Answer: \$LETTER' (without quotes) where LETTER is one of ABCD. Think step by step before answering.

\{\{question\}\}
\end{AIboxC}
\caption{The prompt used in zero shot evaluation prompting}
\label{prompt:zero_shot_cot}
\end{figure}

\begin{figure}[t!]
\centering
\small
\begin{AIboxC}{Direct answer prompting}
Answer the following multiple choice question. Your response should be of the following format: 'Answer: \$LETTER' (without quotes) where LETTER is one of ABCD. Answer immediately, do not think step by step.

\{\{question\}\}
\end{AIboxC}
\caption{The prompt used in zero shot evaluation prompting}
\label{prompt:zero_shot_da}
\end{figure}

\section{Model Finetuning Details}

For all finetuning experiments, we used Low-Rank Adaptation (LoRA) \citep{Hu2021LoRALA} to efficiently adapt the LLaMA 3 8B Instruct model. We set the LoRA rank to 128 and the scaling parameter alpha of 64.

To determine optimal training parameters, we conducted a hyperparameter sweep across three learning rates: 2e-4, 1e-4, and 8e-5. Model selection was performed based on performance on the MMLU validation set. The following hyperparameters were kept constant across all experimental configurations (baseline, supervised finetuning, and DPO):

\begin{description}
    \item Batch size: 16
    \item Maximum sequence length: 4,096
    \item Optimizer: AdamW
    \item Weight decay: 0.1
    \item Learning rate schedule: Cosine decay with 10\% warmup steps
    \item Training epochs: 3
\end{description}

All experiments were conducted on 2 NVIDIA 3090 GPUs with mixed-precision training (BF16). We rely on Axolotl \citep{axolotl2024} for training all models. The total training time for each configuration was approximately 13 hours.
\section{Example Questions which is not suitable to add NA}
\label{app:na_addable_questions}

In the following section we included only partial subjects from each four category due to the large amount of subjects from MMLU. For STEM category we included College Mathematics (Figure \ref{prompt:contain-na-college-mathematics}), Conceptual Physics (Figure \ref{prompt:contain-na-conceptual-physics}) and Astronomy (Figure \ref{prompt:contain-na-astronomy}.

\begin{figure*}[t]
\centering
\begin{AIbox}{College Mathematics}
\paragraph{Suitable to add NA}

Let $T: \mathbb{R}^2 \to \mathbb{R}^2$ be the linear transformation that maps the point $(1, 2)$ to $(2, 3)$ and the point $(-1, 2)$ to $(2, -3)$. Then $T$ maps the point $(2, 1)$ to\\
A. (1, 6)\\
B. (-1, 4)\\
C. (3, 2)\\
D. (-4, 3)\\
\paragraph{Not Suitable to add NA}
Let M be a 5 $\times$ 5 real matrix. Exactly four of the following five conditions on M are equivalent to each other. Which of the five conditions is equivalent to NONE of the other four?\\
A. For any two distinct column vectors u and v of M, the set {u, v} is linearly independent.\\
B. The homogeneous system Mx = 0 has only the trivial solution.\\
C. The system of equations Mx = b has a unique solution for each real 5 $\times$ 1 column vector b.\\
D. The determinant of M is nonzero.
\end{AIbox}
\caption{The second question requires test takers to ignore the missing 5-th conditions which result in a missing condition which the test taker cannot determine , violating rule \#3.}
\label{prompt:contain-na-college-mathematics}
\end{figure*}

\begin{figure*}[t]
\centering
\begin{AIbox}{Conceptual Physics}
\paragraph{Suitable to add NA}
A simple and correct way to comprehend satellites orbiting Earth is to view them as\\
A. balanced between gravitational and centripetal forces.\\
B. beyond the main pull of Earth gravity.\\
C. in mechanical equilibrium with a net force of zero.\\
D. having sufficient tangential velocities to fall around rather than into Earth.\\
\paragraph{Not Suitable to add NA}

The difference between dc and ac in electric circuits is that in dc, charges flow\\
A. steadily in one direction\\
B. in one direction\\
C. to and fro\\
D. All of these
\end{AIbox}
\caption{The second question contains multiple correct answers : one direction in option A and B, violating the rule \#4.}
\label{prompt:contain-na-conceptual-physics}
\end{figure*}

\begin{figure*}[t]
\centering
\begin{AIbox}{Astronomy}
\paragraph{Suitable to add NA}
One astronomical unit (AU) is equal to ...\\
A. 130 million km\\
B. 150 million km\\
C. 170 million km\\
D. 190 million km\\
\paragraph{Not Suitable to add NA}
Why is the Mars Exploration Rover Spirit currently tilted towards the north?\\
A. Because it’s climbing up a big hill.\\
B. Because it’s in the southern hemisphere where it is winter now.\\
C. Because it’s in the northern hemisphere where it is winter now.\\
D. Because one of its wheels broke.

\end{AIbox}
\caption{The question about the Mars Exploration Rover Spirit's tilt involves a specific factual scenario that is not deterministic or based on a finite set of possible answers. The options provided are not exhaustive of all possible reasons for the rover's tilt, and the correct answer (B) is based on a specific situational context rather than a universally verifiable fact.}
\label{prompt:contain-na-astronomy}
\end{figure*}

For Social Science category we include US Foreign Policy (Figure \ref{prompt:no-na-us-foreign-policy}), Econometrics (Figure \ref{prompt:contain-na-econometrics}), High School Geography (Figure \ref{prompt:no-na-high-school-geography})

\begin{figure*}[t]
\centering
\begin{AIbox}{US Foreign Policy}
\paragraph{Suitable to add NA}
'Peace, commerce, and honest friendship with all nations, entangling alliances with none'. Identify the speaker.\\
A. James Madison\\
B. Abraham Lincoln\\
C. Woodrow Wilson\\
D. Thomas Jefferson
\\
\paragraph{Not Suitable to add NA}

How many states in the international system are likely to have nuclear weapons right now?\\
A. Fewer than 7 (Answer)\\
B. Between 8 and 15\\
C. Between 16 and 25\\
D. More than 25
\end{AIbox}
\caption{Reason why the first question is suitable because the question asks for the identification of a speaker of a specific quote, which is a factual question with a clear, verifiable answer. Reason why the second question is not suitable is because the correct answer is based on current geopolitical knowledge, which is not stated when and the answer could change as global powers shift.}
\label{prompt:no-na-us-foreign-policy}
\end{figure*}

\begin{figure*}[t]
\centering
\begin{AIbox}{Econometrics}
\paragraph{Suitable to add NA}
Which of the following statements are true concerning the autocorrelation function (acf) and partial autocorrelation function (pacf)?\\
\\
i) The acf and pacf will always be identical at lag one whatever the model\\
\\
ii) The pacf for an MA(q) model will in general be non-zero beyond lag q\\
\\
iii) The pacf for an AR(p) model will be zero beyond lag p\\
\\
iv) The acf and pacf will be the same at lag two for an MA(1) model\\
A. (ii) and (iv) only\\
B. (i) and (iii) only\\
C. (i), (ii), and (iii) only\\
D. (i), (ii), (iii), and (iv)\\
\paragraph{Not Suitable to add NA}
Which of the following are advantages of the use of panel data over pure cross-sectional or pure time-series modelling?\\
\\
(i) The use of panel data can increase the number of degrees of freedom and therefore the power of tests\\
\\
(ii) The use of panel data allows the average value of the dependent variable to vary either cross-sectionally or over time or both\\
\\
(iii) The use of panel data enables the researcher allows the estimated relationship between the independent and dependent variables to vary either cross-sectionally or over time or both\\
A. (i) only\\
B. (i) and (ii) only\\
C. (ii) only\\
D. (i), (ii), and (iii)
\end{AIbox}
\caption{he reason why the second question is not suitable to add NA is because this is a conceptual question related to econometrics and statistics, which does not have a deterministic or factual, the options provided are not mutually exclusive, and the question does not fit into any of the criteria for replacing the answer with "None of the above." The correct answer, B, is based on understanding the specific advantages of panel data. }
\label{prompt:contain-na-econometrics}
\end{figure*}

\begin{figure*}[t]
\centering
\begin{AIbox}{High School Geography}
\paragraph{Suitable to add NA}
Which pair of religions below share a long-held tradition of cremating their dead?\\
A. Hindu and Buddhist\\
B. Hindu and Muslim\\
C. Muslim and Jewish\\
D. Christian and Buddhist\\
\paragraph{Not Suitable to add NA}
What was the primary reason the Green Revolution did NOT help Africa much?\\
A. It targeted crops Africans don't grow and eat.\\
B. It required sophisticated machinery.\\
C. Population growth has fallen so drastically that Africans don't need the increased food production.\\
D. African agricultural production went up without its help.
\end{AIbox}
\caption{The second question asks about the primary reason the Green Revolution did not help Africa much. This is not a deterministic question with a definitive answer like a math or science question. It is also not a question with a finite set of possible answers, as there could be multiple reasons or interpretations regarding the impact of the Green Revolution on Africa.}
\label{prompt:no-na-high-school-geography}
\end{figure*}

For Humanities we include International Law (Figure \ref{prompt:contain-na-international-law}), High School US History(Figure \ref{prompt:no-na-high-school-us-history}), Jurisprudence (Figure \ref{prompt:no-na-jurisprudence})

\begin{figure*}[t]
\centering
\begin{AIbox}{International Law}
\paragraph{Suitable to add NA}
What are the criteria for statehood under the 1933 Montevideo Convention?\\
A. It requires that the entity in question is not an aggressor and that it is peaceful\\
B. It requires recognition by the majority of other nations\\
C. It requires a permanent population, a defined territory, a government and a capacity to enter into foreign relations\\
D. It requires stable and indissoluble borders as well as recognition\\
\paragraph{Not Suitable to add NA}
What is the meaning of "armed attack" in Article 51 UN Charter?\\
A. Armed attack includes all types of armed force\\
B. Armed attack includes all high intensity instances of armed force (Answer) \\
C. Armed attack includes terrorist attacks\\
D. An "armed attack" gives the right to invade the aggressor State
\end{AIbox}
\caption{The second question is not suited to apply NA because A, B, and C are similar (all about defining armed attack), hence replacing B with NA would resulted in A, C as the correct answers as well.}
\label{prompt:contain-na-international-law}
\end{figure*}

\begin{figure*}[t]
\centering
\begin{AIbox}{High School Us History}
\small
\paragraph{Suitable to add NA}
This question refers to the following information.\\
Here is the case of a woman employed in the manufacturing department of a Broadway house. It stands for a hundred like her own. She averages three dollars a week. Pay is \$1.50 for her room; for breakfast she has a cup of coffee; lunch she cannot afford. One meal a day is her allowance. This woman is young, she is pretty. She has "the world before her." Is it anything less than a miracle if she is guilty of nothing less than the "early and improvident marriage," against which moralists exclaim as one of the prolific causes of the distresses of the poor? Almost any door might seem to offer a welcome escape from such slavery as this. "I feel so much healthier since I got three square meals a day," said a lodger in one of the Girls' Homes. Two young sewing-girls came in seeking domestic service, so that they might get enough to eat. They had been only half-fed for some time, and starvation had driven them to the one door at which the pride of the American-born girl will not permit her to knock, though poverty be the price of her independence.\\
—Jacob Riis, How the Other Half Lives, 1890\\
Riis's work as an investigator of the lives of the poor can most directly be associated with which of the following?\\
A. Yellow Journalism\\
B. Abolitionism\\
C. The muckrakers\\
D. Socialism\\
\paragraph{Not Suitable to add NA}
This question refers to the following information.\\
"The challenge of the next half century is whether we have the wisdom to use wealth to enrich and elevate our national life, and to advance the quality of our American civilization….The Great Society rests on abundance and liberty for all. It demands an end to poverty and racial injustice, to which we are totally committed in our time. But that is just the beginning. The Great Society is a place where every child can find knowledge to enrich his mind and to enlarge his talents. It is a place where leisure is a welcome chance to build and reflect, not a feared cause of boredom and restlessness. It is a place where the city of man serves not only the needs of the body and the demands of commerce but the desire for beauty and the hunger for community. It is a place where man can renew contact with nature. It is a place which honors creation for its own sake and for what it adds to the understanding of the race. It is a place where men are more concerned with the quality of their goals than the quantity of their goods. But most of all, the Great Society is not a safe harbor, a resting place, a final objective, a finished work. It is a challenge constantly renewed, beckoning us toward a destiny where the meaning of our lives matches the marvelous products of our labor."\\
Lyndon Johnson, Remarks at the University of Michigan, Ann Arbor, 1964\\
Which one of the following was an unintended consequence of the liberal successes of the 1960s?\\
A. Liberal Democrats abandoned anti-war protests in a show of support for President Johnson.\\
B. Conservative Republicans mobilized to defend traditional mores and curb government authority.\\
C. Economic recession catalyzed by increased government spending causing "stagflation."\\
D. A majority of Northern black voters abandoned the Democrat party, siding with Republicans.
\end{AIbox}
\caption{The reason why the second questio is not suitable to add NA is because the options provided are not deterministic or factual in the sense of having a single, verifiable answer like a math problem or a historical date.}
\label{prompt:no-na-high-school-us-history}
\end{figure*}

\begin{figure*}[t]
\centering
\begin{AIbox}{Jurisprudence}
\paragraph{Suitable to add NA}
Which of the following statements is correct concerning the "reasonable person" standard in tort law?\\
A. The reasonable person standard varies from person to person.\\
B. The reasonable person standard focuses on the defendant's subjective mental state rather than on the defendant's behavior\\
C. A person with a physical disability must act as would a reasonable person with the same disability.\\
D. A person with a mental disability must act as would a person with the same mental disability.\\
\paragraph{Not Suitable to add NA}
Austin has been described as a 'naive empiricist.' Why?\\
A. Because he neglects the importance of morality.\\
B. Because his account of law is based on an anachronistic model of a legal system.\\
C. Because he conceives of laws in a pragmatic rather than a conceptual manner.\\
D. Because he overlooks the role of law in economic relations.
\end{AIbox}
\caption{The second question is not well suited to add NA because the question is more interpretative and subjective, likely based on philosophical or theoretical analysis, which does not lend itself to a "None of the above" option. The answer choice "C" is based on a specific interpretation of Austin's views, which may not be universally agreed upon or verifiable in the same way as a factual or deterministic question.}
\label{prompt:no-na-jurisprudence}
\end{figure*}

For Other category we include Nutrition(Figure \ref{prompt:no-na-nutrition}), Global Facts(Figure \ref{prompt:no-na-global-facts}), Marketing (Figure \ref{prompt:no-na-marketing})

\begin{figure*}[t]
\centering
\begin{AIbox}{Nutrition}
\paragraph{Suitable to add NA}
Which statement about the oral phase of digestion is INCORRECT?\\
\\
A. About 2\% of the energy content of food is expended during the action of chewing and swallowing it.\\
B. Swallowing involves contraction and relaxation of at least 14 groups of muscles in about 10 seconds in healthy subjects\\
C. The biofilm covering tooth enamel contains several salivary and bacterial enzymes\\
D. Salivary amylase digests the dextran film on tooth enamel formed from dietary sucrose
\paragraph{Not Suitable to add NA}
Which of the following confirmed values meet the diagnostic threshold for diabetes? \\
\\
A. fasting blood glucose ? 140 mg/dl\\
B. random glucose > 160 mg/dl\\
C. 2 hour post prandial glucose $\geq$ to 126 mg/dl\\
D. fasting blood glucose $\geq$ 126 mg/dl (Answer)
\end{AIbox}
\caption{The second question's contain option A which is not well represented and numeric number in C, D contains similar number, which violates rule \#1.}
\label{prompt:no-na-nutrition}
\end{figure*}

\begin{figure*}[t]
\centering
\begin{AIbox}{Global Facts}
\paragraph{Suitable to add NA}
Which of the following countries emitted the most CO2 in 2017?\\
A. Canada\\
B. Russia\\
C. Iran\\
D. Japan\\
\paragraph{Not Suitable to add NA}
At its peak, what was the approximate difference in approval of school desegration from the South and the rest of the U.S.?\\
A. 80\% (Answer)\\
B. 40\%\\
C. -40\%\\
D. -80\%
\end{AIbox}
\caption{Since the second question refers to "approximate" replacing 80\% answer would result in 40\% being the next possible answer in-line.}
\label{prompt:no-na-global-facts}
\end{figure*}

\begin{figure*}[t]
\centering
\begin{AIbox}{Marketing}
\paragraph{Suitable to add NA}
\underline{\hspace{2cm}} can be defined as the aspect of our psyche that determines the way in which 
we respond to our environment in a relatively stable way over time.\\
A. Perception.\\
B. Personality.\\
C. Learning.\\
D. Memory.\\
\paragraph{Not Suitable to add NA}
Second-hand data, collected for someone else's purposes, is known as\underline{\hspace{2cm}}:\\
A. Primary research.\\
B. Descriptive research.\\
C. Causal research.\\
D. Secondary research.
\end{AIbox}
\caption{The second question contain bad options clarity which violates rule \#1, \#2.}
\label{prompt:no-na-marketing}
\end{figure*}

\end{document}